\documentclass[prc,twocolumn,showpacs,preprintnumbers,amsmath,amssymb,nofootinbib]{revtex4-1}

\usepackage{graphicx}
\usepackage{dcolumn}
\usepackage{bm}
\usepackage[mathlines]{lineno}
\usepackage{multirow}
\usepackage{CJK}

\newcommand{\rr} {\boldsymbol{r}}

\begin{document}
\begin{CJK*}{GBK}{song}
\title{ Thermal  fission rates with temperature dependent fission barriers}

\author{Yi Zhu}
\author{J.C. Pei }
\email{peij@pku.edu.cn}
\affiliation{State Key Laboratory of Nuclear
Physics and Technology, School of Physics, Peking University,  Beijing 100871, China}

\begin{abstract}
\begin{description}

\item[Background] The fission processes of thermal excited nuclei are conventionally studied by
statistical models which rely on inputs of phenomenological level densities and potential barriers. Therefore
the microscopic descriptions of spontaneous fission and induced fission are very desirable
 for a unified understanding of various fission processes.

\item[Purpose] We propose to study the fission rates, at both low
and high temperatures,   with microscopically calculated
temperature-dependent fission barriers and collective mass parameters.

\item[Methods] The fission barriers are calculated by the finite-temperature
Skyrme-Hartree-Fock+BCS method. The mass parameters
are calculated by the temperature-dependent cranking approximation. The thermal fission rates can be
obtained by the imaginary free energy approach at all temperatures, in which fission barriers are naturally temperature dependent.
The fission
at low temperatures can be described mainly as a barrier-tunneling process. While the
fission at high temperatures  has to incorporate the reflection above
barriers.

\item[Results] Our results of spontaneous fission rates reasonably agree with
other studies and experiments. The temperature dependencies of fission barrier heights
and curvatures have been discussed. The temperature dependent behaviors
of mass parameters have also been discussed. The thermal fission rates
from low to high temperatures with a smooth connection have been given by different approaches.

\item[Conclusions]  Since the temperature dependencies of fission barrier heights and curvatures, and the mass parameters can
vary rapidly for different nuclei, the microscopic descriptions of thermal fission rates
are very valuable. Our studies without free
parameters  provide a consistent picture to study various fissions such as that in fast-neutron reactors,
 astrophysical environments and fusion reactions for superheavy nuclei.

\end{description}
\end{abstract}

\pacs{}

\maketitle
\end{CJK*}

\section{Introduction}

The microscopic description of the fission process as a large amplitude collective motion is one of the well-known challenges in nuclear many-body theory,
and still large uncertainties exist towards a predictive theory of fission~\cite{Younes08,Schunck16}. Basically, the spontaneous fission can be described as quantum tunneling
based on potential barriers and collective mass parameters, which can be microscopically calculated by nuclear energy density functional theory.
In this respect, there are a number of approaches to describe the collective mass such as the cranking approximation~\cite{Baran11}, the Generate-Coordinate Method (GCM)~\cite{Reinhard1987},
the Adiabatic-Time-Dependent Hartree-Fock-Bogoliubov approach (ATDHFB)~\cite{Baran11}, and the local QRPA method~\cite{nobuo}.
For fission barriers, there are also many efforts either to improve descriptions of potential energy surfaces at large deformations~\cite{markus2} or to seek
multi-dimensional constrained calculations~\cite{zhousg,Staszczak09,Sadhukhan13}.

In addition to issues involved in spontaneous fission, the description of thermal fission in excited nuclei is a more demanding task.
From low to high
 temperatures, the fission process is gradually evolved from the quantum tunneling to the statistical escape mechanism.
 For applications, the thermal fission has a wide range of interests such as the neutron induced fission in reactors and in astrophysical environments, and
 fusion reactions for superheavy nuclei.
 Conventionally, the thermal fission is described by the Bohr-Wheeler transition-state-theory
 and later the dynamical Kramers theory~\cite{Krappes}. The imaginary free energy approach (Im\textsl{F})
 is a general thermodynamic method to calculate thermal quantum decay rates at all temperatures~\cite{langer,Affleck1981},
 which has been widely
 applied to decays of metastable states such as nuclear fissions~\cite{Hagino96} and chemical reactions~\cite{kryvohuz}.
 These methods rely on inputs of barriers or level densities, which are dependent on temperatures, deformations and shell structures.
 As a consequence,
 many corrections and associated parameters have been introduced to interpret experimental results.
Therefore, a consistent description of thermal fission with microscopic inputs that are free of parameters, from low to high temperatures, is very desirable.

In a microscopic view, the thermal excited nuclei can be described  by the finite-temperature Hartree-Fock-Bogoliubov (FT-HFB) theory (or FT-HF+BCS)~\cite{Goodman1981}.
In FT-HFB, the thermal excitations of compound nuclei are described as quasiparticle excitations due to a finite temperature in a heat bath.
The quantum effects: the superfluidity and shell effects, would self-consistently fade away with increasing temperatures~\cite{Egido}.
The fission barriers can be either isothermal or isentropic in terms of free energies~\cite{pei09}.
In previous works~\cite{pei09,pei10,zhuy}, we have studied the neutron emission rates and fission barriers in compound superheavy nuclei microscopically.
We feel an obligation to study further the thermal fission rates with the temperature dependent fission barriers.

In the Kramers and Im\textsl{F} methods, the fission barriers are in terms of free energies which are naturally temperature dependent~\cite{Krappes,Affleck1981}.
It has been realized that the temperature dependent fission barrier should be considered in fission calculations~\cite{newton}.
It is turned out that the Bohr-Wheeler fission calculations also have to introduce
 damping factors to describe the decreasing fission barriers with increasing excitation energies to
 interpret survival probabilities of compound nuclei~\cite{itkis,xia}.
We have demonstrated that the temperature dependence of fission barriers can vary rapidly for specific nuclei~\cite{pei09,Sheikh}, indicating
 non-negligible shell effects in hot-fusion reactions.  This has attracted much attention from experimentalists~\cite{hamilton,loveland}.
 Further, the temperature dependent fission barriers
 in two-dimensional deformation spaces have been studied~\cite{McDonnell}, showing the fission modes become symmetric at high temperatures.
  In addition to the fission barrier heights,
 the curvatures around the equilibrium point and the saddle point can also be
 dependent on temperatures, which are essential inputs for Kramers and Im\textsl{F} methods. This can also
 be microscopically described but has rarely been discussed.

Another essential ingredient for describing thermal fission is the temperature dependent mass parameter.
This has been studied by several phenomenological mean-field methods with the finite-temperature cranking approximation~\cite{Iwamoto1979-1,Baran1981}.
It is difficult to consider the temperature dependence in
 GCM and ATDHFB calculations of mass parameters.  There has also
been microscopic studies with Gogny forces for the temperature dependent cranking mass parameters~\cite{Martin09}.
In fact, the temperature dependent mass parameters have not yet been incorporated into serious calculations of
thermal fission rates.

In this work, we intend to study the thermal fission rates with microscopically calculated
temperature dependent fission barriers and mass parameters. The calculations are based
on the finite-temperature Skyrme Hartree-Fock+BCS framework in deformed coordinate spaces including octupole deformations.
The coordinate-space calculations can naturally describe very elongated nuclear shapes.
The mass parameters are calculated with the cranking approximation with temperatures.
The thermal fission rates  in principle can be described consistently by the Im\textsl{F} method from low  to high temperatures.
At low temperatures, the quantum tunneling process is dominated and the WKB method is adopted.
At high temperatures, the semiclassical reflection process above barriers is considered~\cite{maitra}.
 With microscopic inputs of potential barriers and mass parameters, we will see that the thermal fission from low
to high temperatures can be described in a consistent picture.

Presently our studies are restricted to one-dimensional fission although both quadrupole and octupole deformations are included.
Indeed, the multi-dimensional fission descriptions
are more realistic and computationally more costly.
 In the semiclassical approximation, a multi-dimensional tunneling problem
can be transformed into an effective one-dimensional problem~\cite{kryvohuz}.
Thus the one-dimensional thermal fission has already involved essential
issues in the multi-dimensional fission.
Besides, we have not considered the viscosity and dissipations
which are important at high temperatures. Thus our studies are limited  to a moderate temperature of $T$=1.5 MeV that has
already included the hot-fusion reactions for superheavy nuclei.
For realistic non-adiabatic descriptions,
it is known that the real-time-dependent density functional theory for fission dynamics is only applicable after saddle points~\cite{bulgac,nazarewicz,schunck16,stevenson}, which are useful for studying fission fragment distributions.
In this case the semiclassical descriptions of the thermal fission process with microscopic inputs are promising for multi-dimensional problems~\cite{srihari}.
The present paper can be seen as a basic theoretical attempt
towards fully microscopic descriptions of the thermal fission, instead of the conventional statistical models.

\section{Theoretical framework}\label{theory}

In this section we will discuss the theoretical methods to calculate the spontaneous fission rates,
the temperature dependent fission barriers and mass parameters, and the thermal fission rates.
The thermal fission rates from low to high temperatures are given by the Im\textsl{F} method.

\begin{figure}[t]
  \includegraphics[width=0.48\textwidth]{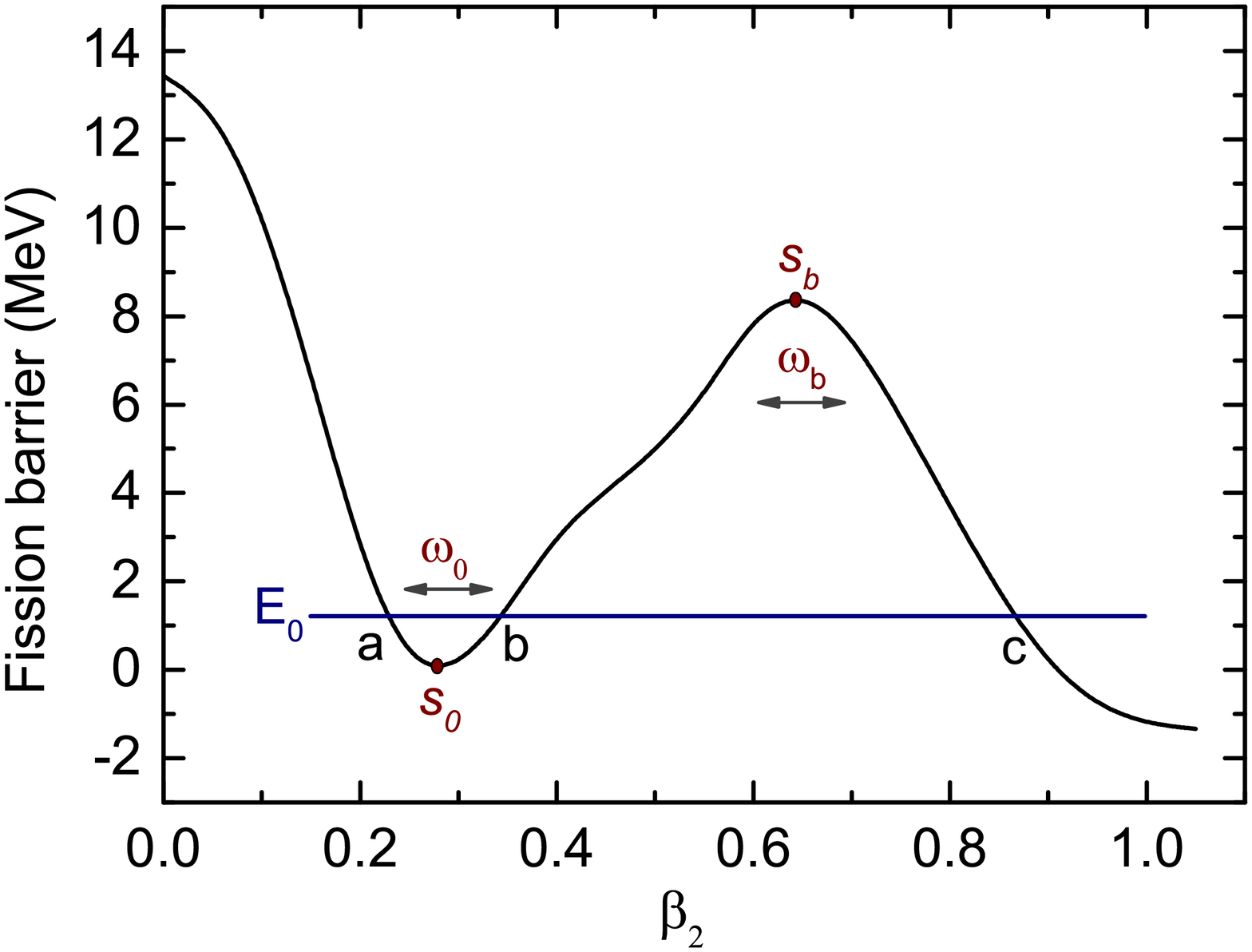}\\
  \caption{(Color online) The fission barrier of $^{260}$Fm is shown to illustrate calculations of fission processes with a decay energy $E_0$.
  The potential frequencies (or curvatures) around the equilibrium point $s_0$
  and the saddle point $s_b$ are labeled as $\omega_0$ and $\omega_b$ respectively.    }
  \label{fig-spon}
\end{figure}

\subsection{Spontaneous fission rates}

The spontaneous fission rates can be  evaluated by the WKB method as a quantum tunneling process along fission pathways~\cite{Erler12,Staszczak13}.
The fission path is obtained by the constrained Skyrme-Hartree-Fock+BCS calculations in the axially-symmetric
coordinate-space, including the reflection asymmetry. The Skyrme-Hartree-Fock+BCS equation is solved by the SKYAX solver~\cite{skyax}.
 In our calculations, the
Skyrme interaction SkM$^{*}$~\cite{skm} and the mixed pairing interaction~\cite{mix-pairing} have been adopted. The SkM$^{*}$ parameter set has
been optimized by including fission barrier heights and has been widely used for microscopic fission studies.
The pairing strengthes are taken as $V_p$=522 and $V_n$=435 MeV fm$^{-3}$ by fitting the pairing gaps of $^{252}$Fm.

The fission width $\Gamma$ along the fission pathway $s$ can be calculated by $\Gamma = P/F$ as~\cite{Erler12},
\begin{subequations}
\begin{equation}
P = \Big[1+\exp\big(2\int_b^c\sqrt{2{M(s)}(V(s)-E_0)}ds\big)\Big]^{-1} ,
\end{equation}
\begin{equation}
F=\int_a^bds \Big(\sqrt{\frac{(E_0-V(s)}{{2M(s)}}}\Big)^{-1} ,
\end{equation}
\label{espon}
\end{subequations}
where the tunneling energy $E_0$, tunneling points $a$, $b$ and $c$ are illustrated in Fig.~\ref{fig-spon}.
The potential energy surface $V(s)$ is given by Skyrme-Hartree-Fock+BCS calculations of binding energies.
The collective mass parameter $M(s)$ is given by the cranking approximation.
The fission lifetime is
calculated by $\hbar/\Gamma$.
In Eq.(\ref{espon}), $P$ is the penetration probability and $F$ is an approximate normalization factor before tunneling, which is similar to the $\alpha$-decay formula~\cite{buck}.
The normalization factor is actually related to the assaulting rate, which has been approximately taken as 10$^{20.38}$ per second in Refs.~\cite{Baran1981,Staszczak13}.

To explain the normalization factor, we assume the potential valley can be described as a one-dimensional harmonic oscillator potential $\frac{1}{2}M\omega_0^2s^2$, then we have
\begin{equation}
F=\displaystyle \int_a^bds \Big(\sqrt{\frac{E-\frac{1}{2}M\omega_0^2s^2}{{2M}}}\Big)^{-1} = \frac{2\pi}{\omega_0} \\
\label{norm2}
\end{equation}
which demonstrated that $1/F$ is related to the assaulting frequency $\omega_0/2\pi$ on the fission barriers, irrespective of the decay energies.
With the assumption of a harmonic potential, the decay energy $E_0$ and $\frac{1}{2}\hbar\omega_0$ (the collective ground state energy) should be equivalent.
The assaulting rate of  10$^{20.38}$ per second is
related to $E_0$=0.5 MeV and $\hbar\omega_0$=1 MeV.
Note that the calculated fission lifetimes are sensitive to $E_0$ and
it is still an issue to determine $E_0$ in the literature.
An assumption of $E_0$ as 0.7$E_{\rm zpe}$ (the zero-point-energy) was successful to reproduce the experimental results~\cite{Staszczak13}.
For realistic potentials,
we can also estimate $E_0$  by the quantization condition~\cite{Baran15}.
For simplicity, our calculations are restricted to one-dimensional barriers, i.e., the fission path (including octupole deformations) is a function of quadrupole deformation $\beta_{20}$.
It should be more realistic to estimate the collective ground state energy $E_0$ with other degrees of freedom in multi-dimensional cases for complex fission pathways.

We calculate the collective mass parameters microscopically for the WKB calculations of spontaneous fission rates.
Based on Skyrme-Hartree-Fock+BCS calculations, the mass parameter $M_{20}(s)$ is calculated by the perturbative cranking approximation as~\cite{Baran11}
\begin{subequations}
\begin{equation}
 M_{20} = \hbar^2 [\mathcal{M}^{(1)}]^{-1}[\mathcal{M}^{(3)}][\mathcal{M}^{(1)}]^{-1}
\end{equation}
\begin{equation}
\mathcal{M}_{ij}^{(K)} = \frac{1}{2} \sum \frac{<0|Q_i|\mu\nu><\mu\nu|Q_j|0>}{(E_{\mu}+E_{\nu})^K}(u_{\mu}v_{\nu}+u_{\nu} v_{\mu})^2
\end{equation}
\end{subequations}
where $v_{\mu}^2$ is the BCS occupation number; $E_{\mu}$ is the BCS quasiparticle energy.  The perturbative cranking approximation of mass parameters
 can substantially overestimate
the fission rates~\cite{Sadhukhan13}, compared to the non-perturbative cranking approximation,
although the perturbative cranking approximation has been widely used~\cite{Staszczak13}.

\subsection{Temperature dependent fission barriers}

Our main objective in this work is to study the thermal fission rates with temperature dependent fission barriers.
We have previously studied the thermal fission barriers of compound superheavy nuclei with the finite-temperature Hartree-Fock-Bogoliubov method~\cite{pei09,Sheikh}.
In this work, based on the Skyrme-Hartree-Fock+BCS solver, we implement the finite-temperature BCS calculations according
to Ref.\cite{Goodman1981}.  With a given temperature $T$, the normal density $\rho$ and the pairing density $\tilde{\rho}$ have to be modified as
\begin{equation}
\begin{array}{l}
\rho_{T}(\rr)= \sum _{i} [v_i^2(1-f_i)+u_i^2f_i]|\phi_i(\rr)|^2 \vspace{5pt} \vspace{5pt}\\
\tilde{\rho}_T(\rr) = \sum _{i} u_i v_i (1-2f_i) |\phi_i(\rr)|^2\\
\end{array}
\end{equation}
where $f_{i}=1/({1+e^{E_{i}/kT}})$ is the temperature dependent distribution factor, $E_i$ is the BCS quasiparticle energy, $k$ is the Boltzmann constant.  Other density functionals
can also be modified similar to the normal density. The particle number conservation equation is modified as:
\begin{equation}
N = 2\sum_{i>0}[v_i^2 +(u_i^2-v_i^2)f_i].
\end{equation}
The entropy is obtained by
\begin{equation}
S = -k\sum_i[f_i \ln f_i + (1- f_i) \ln(1-f_i)].
\end{equation}

Finally the temperature dependent fission barriers are calculated in terms of
 the free energy $F = E(T)-TS$, where $E(T)$ is the intrinsic binding energy. The temperature dependence of
 fission barriers can be related to the melting down of shell effects and has been found to be important to
 explain the experimental survival probabilities~\cite{itkis}.
 In addition to barrier heights, the temperature dependencies of curvatures of the potential valley and the barrier are also essential inputs for fission calculations,
 which are natural results of microscopic calculations. In this work, the beyond mean-field corrections to potential barriers have not been included,
 which are important at the zero temperature~\cite{libert}. The SkM$^{*}$ force~\cite{skm} that includes fission barriers in the fitting procedure could  partially consider
 such effects.

\subsection{Temperature dependent mass parameters}

 The temperature dependent collective mass parameters can be obtained by the cranking approximation with temperatures.
 Compared to expressions at the zero temperature, the pairing occupation numbers have to be
 explicitly modified as~\cite{Iwamoto1979-1,Baran1994},
 \begin{equation}
 \begin{array}{ll}
\mathcal{M}_{ij, T}^{(K)} &=\displaystyle \frac{1}{2}\sum_{\mu\neq \nu} <0|Q_i|\mu\nu><\mu\nu|Q_j|0> \vspace{5pt}\\
              &\displaystyle \Big\{ \frac{(u_\mu u_\nu - v_\mu v_\nu)^2}{(E_\mu-E_\nu)^K}\big[\text{tanh}(\frac{E_\mu}{2kT})-\text{tanh}(\frac{E_\nu}{2kT})\big] \Big\} \vspace{5pt}\\
              &\displaystyle +\frac{1}{2}\sum_{} <0|Q_i|\mu\nu><\mu\nu|Q_j|0> \vspace{5pt}\\
              &\displaystyle \Big\{\frac{(u_\mu v_\nu + u_\nu v_\mu)^2}{(E_\mu+E_\nu)^K}\big[\text{tanh}(\frac{E_\mu}{2kT})+\text{tanh}(\frac{E_\nu}{2kT})\big] \Big\} \vspace{5pt} \\
\end{array}
\end{equation}
We add a smooth factor of 1.0 in the denominator to avoid numerical divergence when two quasiparticle energies are close.
The behaviors of temperature dependent mass parameters have been studied in several earlier publications~\cite{Iwamoto1979-1,Baran1994,Martin09}.

\subsection{Thermal fission rates at low temperatures}

 The microscopic descriptions of fission process at low temperatures
are very interesting to study the induced fission.
The fission at low temperatures can basically be considered as a quantum tunneling process, based on
the temperature dependent fission barriers and mass parameters.
 In contrast to the spontaneous fission,
the thermal fission involves excited states which are distributed statistically in terms of excitation energies.
The excited states with energies of $E_n$ within the potential valley are quasi-stationary and
can be approximately described by the Bohr-Sommerfeld  quantization condition~\cite{Baran15},
\begin{equation}
 \int_a^b ds\sqrt{2M_T(s)[E_n -V_T(s)]} = (n+1/2)\pi
 \label{equz}
\end{equation}
where $V_T$ and $M_T$ are temperature dependent potential barriers and mass parameters.
For the spontaneous fission, we only consider the tunneling energy $E_0$.
For the thermal fission, we need to consider all the eigen-states with $E_n$ lower than barriers.

Based on the spontaneous fission formula, the average thermal fission width at a temperature $T$ ($\beta=1/kT$) is obtained straightforwardly with the Boltzmann distribution,
and is written as,
\begin{subequations}
\begin{equation}
\Gamma(T) =\frac{\sum_n \text{exp}(-\beta E_n)P(E_n)/F(E_n)}{\sum_n \text{exp}(-\beta E_n)}
\end{equation}
\begin{equation}
P(E_n) = \Big[1+ \text{exp}\big(2\int_b^c\sqrt{{2M_T(s)}(V_T(s)-E_n)}ds\big)\Big]^{-1}
\label{e9b}
\end{equation}
\begin{equation}
F(E_n)=\int_a^bds \Big(\sqrt{\frac{(E_n-V_T(s)}{{2M_T(s)}}}\Big)^{-1}
\end{equation}
\label{espon2}
\end{subequations}
In Eq.(\ref{espon2}), the energies $E_n$ of collective quasi-boundary states within the potential valley are obtained from Eq.(\ref{equz}).
Obviously, this formula is only suitable at very low temperatures since $E_n$ are lower than barriers.  In addition,  Eq.(\ref{espon2})
would be problematic if $\omega_0$ is very large and the number of states within the potential valley is not sufficient.

For comparison, we like to introduce
the imaginary free energy method~\cite{langer,Affleck1981,weiss} which is more strict.
In this method the quantity of interest is the free energy of the metastable system.
To obtain the imaginary part of the free energy, it is key to calculate the partition function as a functional integral over the contour~\cite{langer}.
The decay probability is related to the imaginary free energy.
The Im\textsl{F} formula
at low temperatures is given as~\cite{Affleck1981},
\begin{equation}
\begin{array}{l}
\Gamma = \displaystyle \frac{1}{Z_0} \frac{1}{2\pi\hbar}\int_0^{V_b} P(E)\exp(-\beta E))dE \vspace{5pt} \\
Z_0= \displaystyle [2 \sinh(\frac{1}{2}\beta\hbar\omega_0)]^{-1} \\
\end{array}
\label{eimf}
\end{equation}
where $\omega_0$ is the frequency around the equilibrium point of the potential valley; $V_b$ is the barrier height;
$Z_0$ is the partition function.

We see
the expression Eq.(\ref{espon2}) is similar to the Im\textsl{F} formula Eq.(\ref{eimf}) but with an additional normalization factor $F$, or the assaulting frequency.
We refer Eq.(\ref{espon2}) as the low-temperature Boltzmann fission formula.
The difference of the resulted lifetimes between Eq.(\ref{espon2}) and Eq.(\ref{eimf}) is generally within a factor of 5.
The Boltzmann fission formula can  self-consistently consider the
temperature dependence of the assaulting frequency.
It has been discussed that a slowly changed temperature-dependent assaulting frequency should be more reasonable than
the constant  $(2\pi\hbar)^{-1}$ in the Im\textsl{F} theory~\cite{Waxman1985}.

Without dissipation, the estimated transition temperature from quantum tunnelings to
thermal decays is related to $\omega_b$ by~\cite{Affleck1981}
\begin{equation}
T_c=\frac{\hbar\omega_b}{2\pi k }
\end{equation}
For instance, $T_c$ is 0.24 MeV with $\omega_b$=1.5 MeV,
which is a very low temperature. For realistic potentials, we see that the low temperature Im\textsl{F} formula can be applied to temperatures
that are slightly higher than $T_c$ when the above-barrier decay ratio is still small.

\subsection{Thermal fission rates at high temperatures}

The thermal fission rates of compound nuclei at high temperatures are of great interests for productions of superheavy nuclei.
In particular, the  $^{48}$Ca -induced hot fusion experiments have been very successful~\cite{hamilton}. In contrast to
the fission at low temperatures that is mainly a quantum tunneling process, the thermal fission at high temperatures need to consider the reflection
above barriers.

For energies above barriers, the refection can be considered as a tunneling process in the momentum space~\cite{maitra},
which is difficult to be evaluated for complex-shaped barriers. In a special case, the reflection above
a parabolic potential can be analytically obtained. Therefore, we can approximate the temperature-dependent barrier by an inverted
harmonic oscillator potential,
\begin{equation}
V_{\rm  barrier}(s) = V_b -\frac{1}{2}M\omega_b^2(s-s_{\rm b})^2
\end{equation}
where $V_b$ is the barrier height, $M$ is the mass parameter at the saddle point $s_b$, and $\omega_b$ is the barrier curvature (or frequency).

It is crucial to estimate the fission potential valley frequency $\omega_0$ and the barrier frequency $\omega_b$.
Usually the frequencies are given by the second-order derivative of the potential as:
\begin{equation}
\omega_0=\sqrt{\frac{V^{''}(s_0)}{M(s_0)}}, ~~~~~ \omega_b=\sqrt{-\frac{V^{''}(s_b)}{M(s_b)}}
\label{der2}
\end{equation}
However, the microscopic mass parameters are very much dependent on the deformation coordinates, as shown
in Section \ref{results}.
For realistic potential barriers and mass parameters, we
can extract $\omega_0$ and $\omega_b$ approximately by
\begin{equation}
\begin{array}{l}
\omega_0 = \displaystyle \pi E / \int_a^b\sqrt{{2M_T(s)}(E-V_T(s))}ds  \vspace{5pt}\\
\omega_b = \displaystyle \pi (V_b-E) / \int_b^c\sqrt{{2M_T(s)}(V_T(s)-E)}ds \\
\end{array}
\label{eomg}
\end{equation}
which is exact with a harmonic oscillator potential. In principle, results of Eq.(\ref{der2})
and Eq.(\ref{eomg}) should be close. It is turned out that Eq.(\ref{eomg}) is roughly independent of $E$
and is more reliable by avoiding the uncertainties in searching of minimum and saddle points.

The decay rate with an energy $E$ above the inverted harmonic oscillator potential is written as
\begin{equation}
 \frac{1}{2\pi\hbar}\big\{ 1+\exp(-2\pi(E-V_b)/\hbar\omega_b) \big\}^{-1}.
\end{equation}
According to the Im\textsl{F} method for temperatures higher than $T_c$,
the averaged fission rate after integral over $E$ is written as~\cite{Affleck1981}
\begin{equation}
\Gamma = \frac{\omega_b}{2\pi}\frac{\sinh(\frac{1}{2}\beta\hbar\omega_0)}{\sin(\frac{1}{2}\beta\hbar\omega_b)}\exp(-\beta V_b),
\label{espon3}
\end{equation}
which can be related to the Kramers formula at high temperatures~\cite{Affleck1981}:
\begin{equation}
\Gamma_{\rm Kramers} = \frac{\omega_0}{2\pi}\exp(-\beta V_b).
\label{ekramers}
\end{equation}

The Bohr-Wheeler formula should be basically consistent with the Kramers formula without dissipation~\cite{schmidt}.
While the influences
of barrier widths or $\omega_b$ has not been considered in the Bohr-Wheeler formula and the static Kramers formula, which are
the special cases of the Im\textsl{F} method.
Based on the Im\textsl{F} formula Eq.(\ref{espon3}), we see that the fission lifetimes would be increased by decreasing frequencies $\omega_0$ or $\omega_b$ at high temperatures.

In principle, the Im\textsl{F} method works for thermal quantum decays at all temperatures consistently.
The thermal fission at high temperatures involving dissipations and dynamical effects is  complicated
and the Im\textsl{F} method has been extended to dissipative decays~\cite{Hagino96,Waxman1985,weiss}.
The formula of thermal fission rates also becomes complicated considering the deformation dependent mass parameters~\cite{baojd}.
Nevertheless, the microscopic temperature-dependent $\omega_0$, $\omega_b$ and $V_b$ can provide
an opportunity to look for other absent influences.


\begin{figure}[t]
  \includegraphics[width=0.48\textwidth]{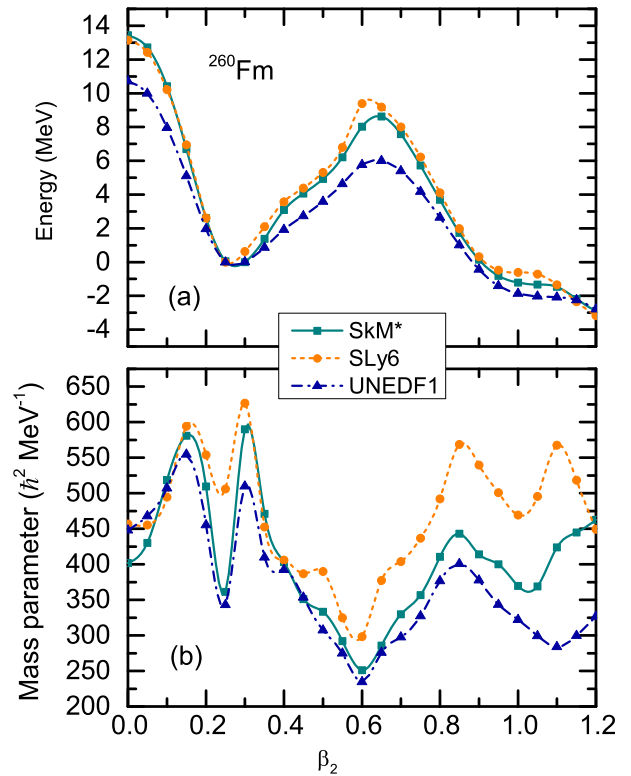}\\
  \caption{(Color online) (a) The calculated spontaneous fission barriers of $^{260}$Fm with different Skyrme forces: SkM$^{*}$, SLy6, and UNEDF1, respectively.
  (b) The calculated collective mass parameters of $^{260}$Fm with the three Skyrme forces.    }
  \label{fig2-spon}
\end{figure}

\begin{figure}[t]
  \includegraphics[width=0.48\textwidth]{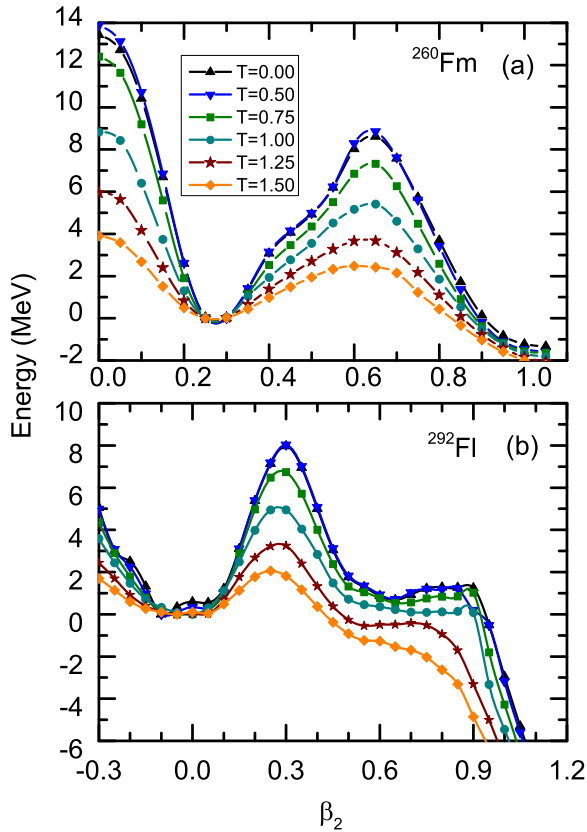}\\
  \caption{(Color online) The calculated temperature dependent fission barriers as a function of
  quadrupole deformation $\beta_2$, (a) for $^{260}$Fm and (b) for $^{292}$114. In $^{292}$114, the reflection
  asymmetric deformation has been taken into account. }
  \label{fig3-tb}
\end{figure}

\section{Results and discussions}\label{results}

In this section we study the spontaneous fission rates and thermal fission rates of some interested nuclei:
$^{240}$Pu, $^{260}$Fm, $^{278}$Cn, and $^{292}$Fl.
$^{240}$Pu has a very long fission lifetime and usually has been chosen for fission benchmark studies~\cite{Schunck16}.
$^{260}$Fm is also an ideal testing case having a single barrier and a primary symmetric fission mode~\cite{Staszczak09}.
$^{278}$Cn and $^{292}$Fl are typical cold-fusion and hot-fusion compound superheavy nuclei in experiments~\cite{hamilton}, respectively.

\subsection{Spontaneous fission rates}

Firstly we studied the spontaneous fission rates of selected nuclei:
$^{240}$Pu, $^{260}$Fm, $^{278}$Cn, $^{292}$Fl, as shown in Table~\ref{table1}.
The calculations are based on the SkM$^{*}$ Skyrme force and the mixed pairing.
It has been pointed out that the cranking mass should be increased to simulate
the ATDHFB mass~\cite{Baran15}. In this work, we adopt the cranking mass that is scaled by a factor of 1.3,
as suggested in Ref.~\cite{Baran15}.
Then $E_0$ is obtained by using the quantization condition.
For $^{240}$Pu, we include the asymmetric fission path of $^{240}$Pu which is important for reducing the
second barrier height. The calculated lifetime is still much
larger than the experimental result mainly due to the absent of non-axial symmetry, which
can reduce the first barrier height~\cite{Burvenich04}.
For $^{260}$Fm,  the calculated fission lifetime agree with that of similar calculations with SkM$^{*}$ in Ref.\cite{Staszczak13}.
$^{292}$Fl has a very long calculated fission lifetime with a small $E_0$, as discussed in the following subsection.
Note that the first barrier and the fission lifetime of $^{292}$Fl could also be reduced by the inclusion of triaxial deformations.
Generally, our results agree with other studies that also adopted the SkM$^{*}$ force.
Indeed, the theoretical lifetimes are expected to be
 reduced with multi-dimensional fission pathways~\cite{Sadhukhan14}.

Note that the fission lifetimes are sensitive to the different approaches to estimate the decay energies $E_0$.
 In this work,
 $E_0$ is related to the potential frequency at the ground state by the Bohr-Sommerfeld quantization condition and is not a free parameter, as given in Table~\ref{table1}.
 Since the potential valley is not a perfect harmonic potential, we keep in mind that the estimation
 of $E_0$ can have considerable uncertainties.
 For example, $E_0$ has to be 1.41 MeV to reproduce the fission lifetime of $^{240}$Pu, which
can reduce the lifetime by 4 orders of magnitude compared to $E_0$=0.92 MeV.
 For $^{292}$Fl, the
 ground state is slightly oblate and has a very soft potential energy surface (shown in Fig.~\ref{fig3-tb}) and the resulted $E_0$ is very small, which
 can substantially increase the fission lifetime.

Fig.\ref{fig2-spon} displays the calculated fission barriers and mass parameters of $^{260}$Fm by three different
Skyrme forces: SkM$^{*}$~\cite{skm}, SLy6~\cite{sly6} and UNEDF1~\cite{markus2}, respectively.  The SkM$^{*}$ and UNEDF1 forces
have been optimized by including fission barrier heights. SLy6 is suitable for
large deformations and surface properties by considering self-consistent center-of-mass corrections~\cite{bender00}.
 One can see that fission barriers of SkM$^{*}$ and SLy6 calculations are close.
On the other hand, the cranking mass parameters of SkM$^{*}$ and UNEDF1 calculations are close.
We note that the small differences in barriers or mass parameters can remarkably
affect the fission rates. Such dependencies can be reduced with minimum action fission pathways
in multi-dimensional calculations~\cite{robledo2}.
The SkM$^{*}$ force has been widely used
for spontaneous fission calculations and is adopted for studies of thermal fission rates in this work.
In addition to the dependence of Skyrme forces, the spontaneous fission lifetimes can also be
reduced significantly with enhanced pairing strengthes, as discussed in Refs.~\cite{Sadhukhan13,Schunck15}.

\renewcommand{\arraystretch}{1.3}
\begin{table}[htb]
 \caption{\label{table1} The calculated spontaneous fission lifetimes (in seconds) of selected nuclei, in which
 $E_0$ is obtained by the quantization condition.
 The experimental data  are also given for comparison.
 }
\begin{ruledtabular}
\begin{tabular}{cccccc}
Nuclei    & Expt (s)~\cite{nndc}       &  $ T_{SF}$ (s)          & $E_0$(MeV)                              \vspace{5pt}   \\  \hline

$^{240}$Pu &  $3.6\times10^{18}$  &  2.73$\times10^{22}$    & 0.92                                                                     \vspace{3pt}     \\
$^{260}$Fm &  $  5.8\times10^{-3} $ &   4.25$\times10^{-3} $  & 0.65                       \vspace{3pt}     \\
$^{278}$Cn&                      &   6.39$\times10^{-5}$  &  0.90                         \vspace{3pt}    \\
$^{292}$Fl&                      &    8.56$\times10^4$   &  0.46
\end{tabular}
\end{ruledtabular}
\end{table}

\begin{figure}[t]
  \includegraphics[width=0.48\textwidth]{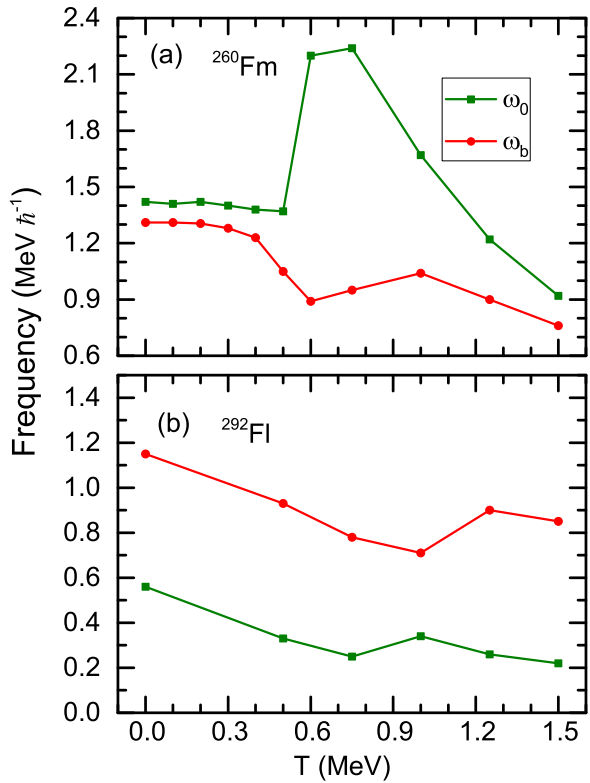}\\
  \caption{(Color online) The calculated potential curvatures (or frequencies) around the equilibrium point ($\omega_0$) and the barrier saddle point ($\omega_b$)
  as a function of temperature, (a) for $^{260}$Fm and (b) for $^{292}$Fl. }
  \label{fig4-omg}
\end{figure}

\begin{figure}[t]
  \includegraphics[width=0.48\textwidth]{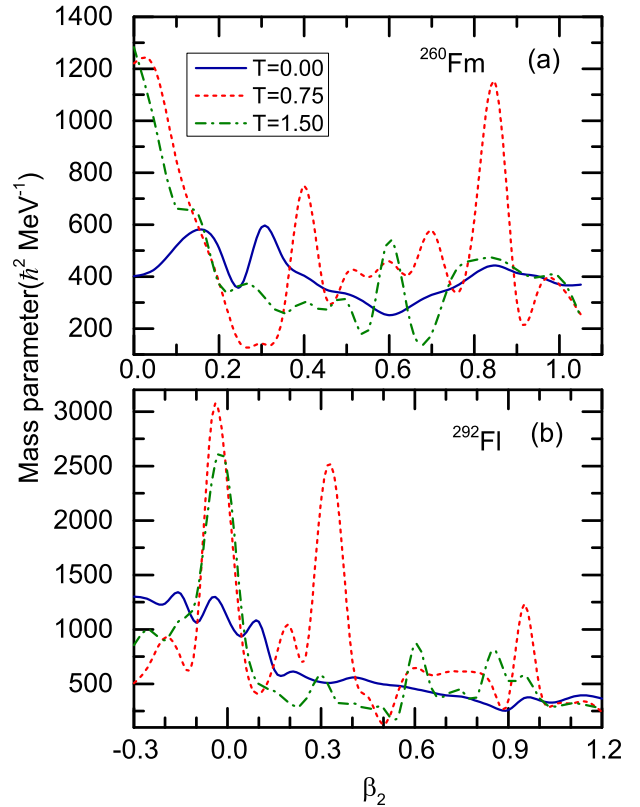}\\
  \caption{(Color online) (a) The calculated temperature-dependent mass parameters as a function of deformations, (a) for $^{260}$Fm and (b) for $^{292}$114. .    }
  \label{fig5-spon}
\end{figure}

\subsection{Temperature dependent fission barriers and mass parameters}

We studied the temperature dependence of fission barriers of selected nuclei, $^{240}$Pu, $^{260}$Fm, $^{278}$Cn and $^{292}$Fl.
Our results obtained with FT-HF+BCS are very close to the earlier FT-HFB results. For example, thermal fission barriers of $^{240}$Pu
has been given in Ref.~\cite{Schunck15}; $^{278}$Cn and $^{292}$Fl have been shown in Ref.~\cite{pei10}. Fig.\ref{fig3-tb} shows the temperature dependent fission barriers
of $^{260}$Fm and $^{292}$Fl. Previously, the asymmetric fission mode of $^{292}$Fl has not been included~\cite{pei10}. In this work,
we do see the asymmetric fission mode is important for $^{292}$Fl and the second barrier is almost gone.
We see the fission barriers are almost unchanged at low temperatures and even slightly increased at $T$=0.5 MeV when the pairing is significantly reduced.
This has also been discussed in several earlier works~\cite{pei10,Schunck15}. After $T$=0.5, the fission barrier heights decrease with increasing temperatures, which
can be described by a damping factor to describe the melting of shell effects~\cite{pei09}.
It is known that microscopic damping factors change rapidly in various nuclei~\cite{pei09,Sheikh},
which are beyond phenomenological descriptions.

In addition to fission barrier heights, the temperature dependent curvatures (or frequencies) around the equilibrium point and the saddle point
are also important.   In Fig.\ref{fig4-omg}, the obtained potential frequencies of $^{260}$Fm and $^{292}$Fl are shown, which are
 estimated by Eq.(\ref{eomg}).
 At temperatures below $T$=0.5 MeV, the frequencies change very slowly.
For  $^{260}$Fm,  it can be
seen that the frequency $\omega_0$ at the equilibrium point increases rapidly  close to $T$=0.6 MeV that is around the pairing phase transition temperature.
$^{292}$Fl is very special with a very small $\omega_0$ that is associated with a very soft equilibrium deformation.
Generally the frequencies $\omega_0$ would be decreased as temperatures increased and compound nuclei would finally become spherical.
The related collective energies $E_n$ would be reduced with increasing temperatures.
For both $^{260}$Fm and $^{292}$Fl, the frequencies $\omega_b$ at the saddle points also decrease as temperatures increase.
Therefore the fission lifetimes can be enhanced due to the decreasing frequencies $\omega_0$ and $\omega_b$ according to the Im\textsl{F} formula at high temperatures.
We see the temperature dependencies of frequencies are different in various nuclei.
This again demonstrated that microscopic calculations of temperature dependent fission barriers are valuable.

Fig.\ref{fig5-spon} shows the temperature dependent behaviors
of mass parameters of $^{260}$Fm and $^{292}$Fl. We studied the temperature dependence of mass parameters of selected nuclei with
the temperature dependent cranking approximation.
Compared to fission barriers, the mass parameters at high temperatures are rather non-smooth.
At zero temperature, the mass parameters is smooth due to the existence of pairing correlations.
At the temperature of $T$=0.75 MeV, it is around the critical temperature for the pairing phase transition and
the mass parameters are increased and become very much irregular. It is known that the
collective inertia mass is inversely proportional to the square of the
pairing gap~\cite{bertsch}. As the temperature
increases from $T$=0 to $T$=0.75 MeV the pairing gap decreases and
therefore the mass parameters must increase.  At the high temperature of $T$=1.5 MeV,
the mass parameters are much reduced and large peaks fade away due to statistical effects.
This behavior has also been shown in Ref.~\cite{Martin09}.
In both $^{260}$Fm and $^{292}$Fl, the mass parameters at spherical shapes increase significantly compared to other deformations.

\subsection{Thermal fission rates from low to high temperatures}

\begin{table}[t]

 \caption{\label{table2} The calculated fission lifetimes of $^{260}$Fm and $^{240}$Pu  at low temperatures, based on the low-temperature Im\textsl{F} approach (see Eq.(\ref{eimf})).
 The corresponding excitation energies are also given in MeV.}
\begin{ruledtabular}
\begin{tabular}{ccccc}
T   &     \multicolumn{2}{c}{$^{260}$Fm}       &  \multicolumn{2}{c}{$^{240}$Pu}               \\
\cline{2-3}\cline{4-5}
(MeV) & E$^*$ & $ T_f$(s) &   E$^*$  & $ T_f$(s) \\  \hline
0.1 &  0.001& 1.50$\times10^{-3}$    &  0.002  &  2.55$\times10^{10}$                \\
0.2 &  0.11 & 1.59$\times10^{-6}$    &  0.13   &  2.80$\times10^{-3}$                 \\
0.3 &  0.83 & 3.67$\times10^{-10}$   &  0.81   &  4.50$\times10^{-8}$                 \\
0.4 &  2.67 & 1.94$\times10^{-12}$   &  2.43   &  3.48$\times10^{-10}$                 \\
0.5 &  5.67 & 7.87$\times10^{-14}$   &  4.85   &  9.08$\times10^{-11}$                 \\
0.6 &  8.63 & 3.48$\times10^{-15}$   &  7.02   &  8.17$\times10^{-12}$                  \\
0.75& 10.91 & 2.07$\times10^{-16}$   &  11.19  &  9.61$\times10^{-13}$                 \\
\end{tabular}
\end{ruledtabular}
\end{table}

\begin{figure}[t]
  \includegraphics[width=0.45\textwidth]{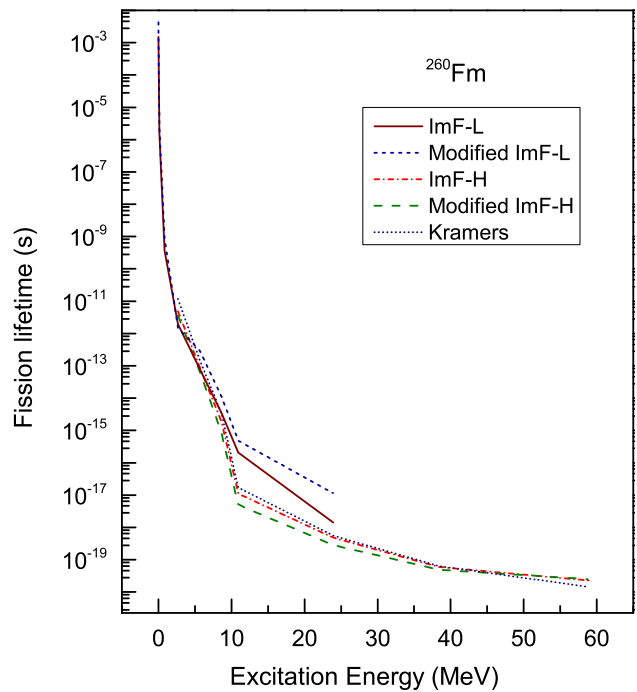}\\
  \caption{(Color online) The calculated thermal fission lifetimes of $^{260}$Fm as a function of excitation energies by different formulas, in which
  `Im\textsl{F}-L' denotes the low-temperature Im\textsl{F} formula Eq.(\ref{eimf}), `Im\textsl{F}-\textsl{H}' denotes the high-temperature Im\textsl{F} formula Eq.(\ref{espon3}),
  `Boltzmann-L' denotes the low-temperature Boltzmann thermal fission formula Eq.(\ref{espon2}),  `Kramers' denotes the Kramers fission formula Eq.(\ref{ekramers}).  }
  \label{fig6-spon}
\end{figure}

In Table \ref{table2}, we studied the temperature dependence of fission rates of $^{240}$Pu and  $^{260}$Fm according
to the low-temperature Im\textsl{F} formula Eq.(\ref{eimf}), from $T$=0.1 to 0.75 MeV. We can see that the calculated fission lifetimes
decrease very rapidly with increasing temperatures. For example, the lifetime has been
decreased by 3 orders in $^{260}$Fm
at an excitation energy of 100 keV (around the astrophysical temperature $T_9$).
At an excitation energy around 5 MeV, its lifetime has been
decreased by 10 orders,  compared to the spontaneous fission lifetime.
The calculated fission lifetimes of $^{240}$Pu decrease even faster than that of $^{260}$Fm.
For $^{240}$Pu, it has a very large $\omega_0$ of 1.87 MeV and a very small $\omega_b$ of 0.54 MeV.
Therefore it has a very low transition temperature, $T_c$=$\omega_b/2\pi$=0.08 MeV, from quantum tunnelings to thermal decays.
While the transition temperature in $^{260}$Fm is $T_c$=0.21 MeV that is much higher than that of $^{240}$Pu.
The low-temperature Im\textsl{F} formula maybe not suitable for $^{240}$Pu  due to a very low $T_c$.
Besides, the fission rates
should be modified considering the double-humped barrier in $^{240}$Pu.
There are a few measurements of thermal fission rates directly~\cite{jacquet}.
Actually it can be related to the fast neutron induced fission cross sections with abundant experimental data.

\begin{table}[htb]
 \caption{\label{table3} The fission lifetimes of selected nuclei are calculated according to the Im\textsl{F} formula at high temperatures (see Eq.(\ref{espon3})).
 The excitation energy and lifetime are given in MeV and seconds, respectively. }
\begin{ruledtabular}
\begin{tabular}{ccccc}
T   &     \multicolumn{2}{c}{$^{260}$Fm}       &  \multicolumn{2}{c}{$^{240}$Pu}               \\
\cline{2-3}\cline{4-5}
(MeV) & E$^*$ & $ T_f$(s) &   E$^*$  & $ T_f$(s) \\  \hline
0.1 &   &     &  0.002   &  4.06$\times10^{16}$                 \\
0.2 &   &    &  0.13   &  4.39$\times10^{-2}$                 \\
0.3 &  0.83 & 1.90$\times10^{-9}$    &  0.81   &  3.25$\times10^{-8}$                 \\
0.4 &  2.67 & 4.90$\times10^{-12}$   &  2.43   &  2.92$\times10^{-11}$                 \\
0.5 &  5.67 & 9.03$\times10^{-14}$   &  4.85   &  4.51$\times10^{-13}$                 \\
0.6 &  8.63 & 1.85$\times10^{-15}$   &  7.02   &  5.51$\times10^{-15}$                  \\
0.75& 10.91 & 1.11$\times10^{-17}$   &  11.19  &   8.13$\times10^{-17}$                 \\
1.0&  23.92 & 4.72$\times10^{-19}$   &  21.22  &   1.12$\times10^{-18}$                 \\
1.25& 38.38 & 6.01$\times10^{-20}$   &  35.42  &   9.14$\times10^{-20}$                 \\
1.5&  58.80 & 2.29$\times10^{-20}$   &  54.40  &   3.27$\times10^{-20}$                 \\
\hline
T   &     \multicolumn{2}{c}{$^{278}$Cn}       &  \multicolumn{2}{c}{$^{292}$Fl}               \\
\cline{2-3}\cline{4-5}
(MeV) & E$^*$ & $ T_f$(s) &   E$^*$  & $ T_f$(s) \\  \hline
0.5 &  4.70   & 3.54$\times10^{-17}$ &  5.82   &  1.01$\times10^{-13}$                 \\
0.75& 11.25   & 3.56$\times10^{-19}$ & 14.1    &  1.25$\times10^{-16}$                  \\
1.0 & 23.17   & 2.32$\times10^{-20}$ &  24.27  &  1.66$\times10^{-18}$                 \\
1.25&  40.17  &                      &  40.22  &  2.09$\times10^{-19}$                 \\
1.5 &  62.34  &                      &  69.01  &  7.33$\times10^{-20}$                 \\
\end{tabular}
\end{ruledtabular}
\end{table}

In Table \ref{table3}, we studied the temperature dependence of the fission rates of selected nuclei
according to the high-temperature Im\textsl{F} formula, which is applicable for $T>T_c$.
Generally, the calculated fission lifetimes at high temperatures decrease less rapidly compared to
the low-temperature rates. The fission rates of $^{240}$Pu and $^{260}$Fm at low temperatures are also given.  In $^{260}$Fm, We indeed see
a smooth connection (or crossover) between low and high temperature formulas at temperatures slightly higher than $T_c$.
For $^{240}$Pu with $T_c$=0.08 MeV, the high temperature Im\textsl{F} formula should be more reasonable from $T$=0.1 MeV,
compared to Table \ref{table2}.
The low temperature Im\textsl{F} formula underestimates the fission lifetimes of $^{240}$Pu at low temperatures and
overestimate fission lifetimes at high temperatures, compared to the high temperature Im\textsl{F} formula.

We see the fission lifetime of {$^{278}$Cn} is smaller than that of {$^{292}$Fl} at high excitation energies by 2 orders.
While such a difference is about 9 orders at zero temperature in Table \ref{table1}. The differences in fission lifetimes of
different nuclei decrease with increasing temperatures as quantum effects fad away.
At $T$=1.25 MeV, the fission barrier of  {$^{278}$Cn} is almost gone in contrast to {$^{292}$Fl}.
The frequency $\omega_0$ in $^{292}$Fl is small that can enhance thermal fission lifetimes.
At $T$=1.0 and 1.5 MeV, its microscopic neutron emission lifetimes~\cite{zhuy} are 1.8$\times10^{-19}$ and 1.7$\times10^{-20}$ seconds,
which are much smaller than its corresponding fission lifetimes of 1.67$\times10^{-18}$ and 7.3$\times10^{-20}$ seconds.
This leads to considerable survival probabilities of {$^{292}$Fl} at high excitations by microscopic calculations,
which are 90$\%$ at $T$=1.0 MeV and 81$\%$ at $T$=1.5 MeV, respectively.

Fig.\ref{fig6-spon} displays the thermal fission lifetimes of {$^{260}$Fm} from low to high temperatures obtained
by different approaches with the same microscopic inputs. Generally the fission lifetimes decease very rapidly at low temperatures and
 decrease slowly at high temperatures.
We see the fission lifetimes by Im\textsl{F} and Kramers formulas are close at high temperatures.
At low temperatures, the Kramers formula overestimates the fission lifetimes.
The Boltzmann fission lifetimes are close to the Im\textsl{F} results at low temperatures.
The fission lifetimes are mainly determined by the barrier heights in the exponential function at high temperatures.
Basically the low and high temperature Im\textsl{F} formulas are consistent although they have different temperature regimes of applicability regarding the transition temperature $T_c$.
The two calculations have comparable results between $T$=0.3 to 0.6 MeV, indicating a smooth transition from quantum tunneling to thermal decays.
After $T$=0.6 MeV, the above-barrier fission is important and the low-temperature formula overestimates the fission lifetimes.
Based on results of $^{260}$Fm,  we see the low-temperature formula can be applied to temperatures that are slightly higher than $T_c$.
In realistic calculations, the crossover of low and high temperature Im\textsl{F} formulas depends on not only $T_c$ (or $\omega_b$) but also the temperature dependent $\omega_0$ and barrier heights.
At temperatures higher than $T$=1.5 MeV, the barriers and quantum effects are almost disappeared and the microscopic calculations would be questionable.

\section{Summary}

In summary, we studied the thermal fission rates with microscopic calculated temperature dependent fission barriers and mass parameters.
The fission lifetime calculations are based on the imaginary free energy method from low to high temperatures in a consistent picture.

In Kramers and Im\textsl{F} methods, the fission barriers are given in terms of free energies which are naturally temperature dependent.
Our calculations involve only the effective Skyrme forces and pairing interactions without free parameters.  We discussed the
temperature dependent behaviors of fission barriers and mass parameters, which change rapidly in various nuclei and are beyond phenomenological descriptions.
Therefore calculations of thermal fission rates with microscopic inputs are very necessary. With the previous microscopic neutron emission rates,
we obtained considerable survival probabilities of {$^{292}$Fl} at high excitations.
We also emphasized the role of potential curvatures $\omega_0$ and $\omega_b$ in the Im\textsl{F} formula.
The curvatures are slowly decreasing from microscopic calculations  at high temperatures and can enhance fission lifetimes.
As a complementary, the spontaneous fission rates have also been studied.
Our studies
can be very useful for microscopic understandings of induced fission in reactors and the astrophysical $r$-process, and survival probabilities of compound superheavy nuclei.
We noticed that large uncertainties still exist towards fully microscopic descriptions of thermal fission rates.  In the future, it is worth to study
both thermal fission rates and fragment distributions by semiclassical methods with microscopic inputs in multi-dimensional spaces.

\section{acknowledgments}
Discussions with  W. Nazarewicz, P. Ring, J. Sadhukhan, N. Schunck  and F.R. Xu are gratefully
acknowledged.
This work was supported by the Research Fund for
the Doctoral Program of Higher Education of China (Grant
No. 20130001110001), and the National Natural Science Foundation of China under Grants No.11375016, 11522538, 11235001.


\end{document}